\documentclass{emulateapj}

\newcommand{\htwo}{H\,{\scriptsize\sc{II}} }

\begin{document}

 \title{Subarcsecond Imaging of the High-Mass Star Forming Region Onsala~1}

 \author{Yu-Nung Su, Sheng-Yuan Liu, and Jeremy Lim}

 \affil{Institute of Astronomy and Astrophysics, Academia Sinica, P.O. Box 23-141, Taipei 106, Taiwan; ynsu@asiaa.sinica.edu.tw}

\begin{abstract}
We report subarcsecond images of the high-mass star forming region
Onsala~1 (ON~1) made with the Submillimeter Array (SMA) at 0.85 mm
and the Very Large Array at 1.3 cm and 3.6 cm.  ON~1 is one of the
smallest ultracompact \htwo regions in the Galaxy and exhibits
various star formation signposts. With our VLA and SMA observations,
two new cm-wave sources and five sub-mm dust sources, respectively,
within a field of $\sim$5$\arcsec$ (corresponding to a linear scale
of 0.05 pc) are identified, indicating the multiplicity at the
center of the ON~1 region. The dust and gas masses of these sub-mm
sources are in the range of 0.8 to 6.4 \emph{M}$_\sun$. Among the
five sub-mm dust sources, SMA2, with a dust and gas mass of 2.6
\emph{M}$_\sun$, demonstrates several star formation signatures, and
hence likely represents an intermediate-mass (or even high-mass)
star forming core. Due to the low star formation efficiency of
$\sim$10\%, we suggest that star formation in the ON~1 region will
continue. For example, SMA4 and SMA5 are not associated with any
star formation signatures and likely mark star formation cores at
very early evolutionary stages.

\end{abstract}


\section{Introduction \label{intro}}

Onsala 1 (hereafter ON~1, also known as G69.54$-$0.98) is recognized
as one of the most compact and isolated \htwo regions in the Milky
Way \citep{zhe85}. Measurements of the kinematic distance to ON~1
span a wide range, from approximately 1 to 6 kpc. We adopt here a
near kinematic distance of 1.8 kpc favored by most studies
\citep[e.g.,][]{mac98,kum04,nag07}. At this distance, the
ultracompact (UC) \htwo region ON~1 has a diameter of
$\sim$1000$-$2000 AU only \citep{kur94}. It is associated with the
luminous ($\sim$10$^4$ \emph{L}$_\odot$) far-infrared source IRAS
20081+3122 \citep{mac98}, and also surrounded by an embedded young
cluster identified in near-IR emission \citep{kum03}. Single dish
observations in CS (5$-$4) and 0.35 mm continuum detected a massive
(about a few $\times$ 100 \emph{M}$_\sun$ scaled to the distance of
1.8 kpc) star forming clump of a size of $\sim$0.5 pc roughly
centered at the position of the UC \htwo region ON~1
\citep{shi03,mul02}. Toward the center of the ON~1 region, various
H$_2$O, OH, and CH$_3$OH maser transitions are detected
\citep[e.g.,][]{dow79,ho83,arg00,nam06,kur04}. To have a consistent
nomenclature, in the paper, we use ``UC \htwo region ON~1'' or
``ON~1'' for short to refer to the \htwo region and ``ON~1 region''
to refer to the cloud core associated with the \htwo region.

High-resolution observations provide clues of multiplicity of
massive stars at the center of the ON~1 region. Observations at 0.85
mm with higher angular resolution ($\sim$3$\arcsec$) resolved a
second dusty clump located approximately 1.5$\arcsec$ to the
northeast of the UC \htwo region ON~1 \citep{su04}, although
observations of 3 mm continuum at 5$\arcsec$ resolution revealed
only a compact source centered on ON~1. Furthermore,
\citet{kum02,kum04} interpreted near-IR H$_2$ and millimeter CO and
H$^{13}$CO$^+$ features as multiple outflows in this region: one
along the east-west direction revealed in the near-IR H$_2$ emission
and CO (2$-$1) transition and another along a position angle of
$\sim$44$^\circ$ traced by the H$^{13}$CO$^+$ (1$-$0) emission. The
driving sources of both outflows are likely located in the close
vicinity of ON~1.

We report here the results of interferometric observations with the
Very Large Array\footnote{The National Radio Astronomy Observatory
is operated by Associated Universities, Inc., under cooperative
agreement with the National Science Foundation.} (VLA) and the
SubMillimeter Array\footnote{The Submillimeter Array is a joint
project between the Smithsonian Astrophysical Observatory and the
Academia Sinica Institute of Astronomy and Astrophysics, and is
funded by the Smithsonian Institution and the Academia Sinica.}
(SMA) in the continuum and molecular lines at sub-arcsecond
resolutions. First we conducted high-resolution SMA observations to
understand the nature of the continuum source adjacent to the ON~1
identified by our previous SMA observations at lower resolutions
\citep{su04}, and a couple of sub-mm continuum sources were
revealed. We then carried out VLA observations to further study
these sub-mm sources. Using the multi-wavelength high-resolution
continuum and molecular-line images, together with the available
results, we explore the nature of the newly identified continuum
sources, investigate clustered mode of star formation in this
region, and discuss the observed chemical variations. We compare the
very accurate maser positions obtained from recent interferometric
maser observations \citep[e.g.,][]{fis07a,nam06,nag07,tre08} with
other signposts of star-formation activities, and further clarify
the relationship among different tracers.

\begin{deluxetable*}{lccc}
\singlespace
 \tabletypesize{\footnotesize} \tablecaption{\sc Onsala 1 Observational Parameters}
 \tablewidth{0pt}
 \tablehead{\colhead{Parameter} & \colhead{VLA 8.4/22.4 GHz} & \colhead{SMA 345 GHz} &
 \colhead{}}
 \startdata
 Primary beam HPBW                          &  $\sim$5.4$\arcmin$/2$\arcmin$        &  $\sim$35$\arcsec$                       & \\
 Phase center~~$\alpha$(\emph{J}2000)       & 20$^h$10$^m$09.08$^s$                 &  20$^h$10$^m$09.08$^s$           & \\
 ~~~~~~~~~~~~~~~~~~~$\delta$(\emph{J}2000)  & 31$^\circ$31\arcmin35.70$\arcsec$     &  31$^\circ$31\arcmin35.70\arcsec & \\
 Projected baseline range (m)               & $\sim$240$-$11000                     & 25$-$220                         &\\
 Cannel spacing (MHz/km s$^{-1}$)           &    ---                                &  0.8125/0.7                       &   \\
 rms noise level (mJy beam$^{-1}$)          &                                       &                                   & \\
                ~~~~continuum               &      0.05/0.06                        &  3$^a$                           & \\
                ~~~~line                    &    ---                                & 100                               & \\
 synthesized beam (arcsec)                  & 0.69$\times$0.60/0.35$\times$0.26     & 0.74$\times$0.54$^b$/0.80$\times$0.50$^c$  & \\
 Gain calibrator                            & 2023+318                              &  BL Lac                          & \\
 Flux calibrator                            & 3C48                                  & Uranus                           & \\
 Passband calibrator                        & ---                                   & 3C279                            & \\
\enddata
 \tablenotetext{a}{the data presented in \citet{su04} incorporated}
 \tablenotetext{b}{for continuum channel, the data presented in \citet{su04} incorporated}
 \tablenotetext{c}{for line channel \label{tb-obs}}
\end{deluxetable*}

\section{Observations and Data Reduction  \label{obs}}

The 345 GHz (0.85~mm) observations were carried out with the SMA on
2004 August 28 in its extended configuration with seven antennas in
the array. See \citet{ho04} for more complete specifications of the
SMA. The observational parameters are summarized in
Table~\ref{tb-obs}. The phase reference center of the observations
was the position of the UC \htwo region ON~1 reported by
\citet{kur04}. The half-power width of the SMA primary beam at 0.85
mm was $\sim$35$\arcsec$. We tuned the receivers to the CS (7$-$6)
line at 342.8830 GHz in the lower sideband (LSB) and the HCN (4$-$3)
line at 354.5055 GHz in the upper sideband (USB).  Note that the 2
GHz bandwidth in each sideband allowed for CH$_3$OH
(13$_1$A$^-$$-$13$_0$A$^+$) at 342.7298 GHz and SO (8$_8$$-$7$_7$)
at 344.3106 GHz to be observed simultaneously. The spectral
resolution was 0.8125 MHz, corresponding to a velocity resolution of
$\sim$0.7 km s$^{-1}$. We smoothed our data to 1.0 km s$^{-1}$
resolution for the analysis presented below. The absolute flux
density scale was determined from observations of Uranus, and is
estimated to have an uncertainty of $\sim$25\%. The quasars 3C279
and BL Lac were observed for bandpass and gain calibration
respectively. We calibrated the data using the MIR software package
adapted for the SMA from the MMA software package developed
originally for the OVRO \citep{sco93}. We made maps using the MIRIAD
package. The rms noise level in a 1.0 km s$^{-1}$ velocity bin was
$\sim$100 mJy beam$^{-1}$. We incorporated the data presented in
\citet{su04} (made with the SMA in its compact configuration) to
make a continuum map with a noise level of $\sim$3 mJy beam$^{-1}$.
The systemic LSR velocity of the ON~1 region is 12 km s$^{-1}$.

Radio continuum observations at 8.4 and 22.4 GHz (i.e., 3.6 and 1.3
cm) were carried out with the VLA on 2005 March 25 in its B
configuration. The pointing center is the same as that of the SMA
observations, and the primary beam is approximately 5.4$\arcmin$ and
2$\arcmin$ at 8.4 and 22.4 GHz respectively. The observational
parameters are also summarized in Table~\ref{tb-obs}. At both
frequencies, the quasars 3C48 and 2023+318 were observed as the flux
and complex gain calibrator respectively. We calibrated the data
using the AIPS package. Self-calibrations were performed in order to
improve the dynamic range of the images to $\gtrsim$400. With
uniform weighting, the angular resolutions at 8.4 and 22.4 GHz were
0.69$\arcsec$ $\times$ 0.60$\arcsec$ at P.A. of $-$51.6$^\circ$ and
0.35$\arcsec$ $\times$ 0.26$\arcsec$ at P.A. of $-$57.8$^\circ$
respectively. The noise levels of the 8.4 and 22.4 GHz continuum
maps were approximately 0.05 and 0.06 mJy beam$^{-1}$ respectively.

As described in \S\ref{discuss-cs}, our SMA observations do not
detected CS (7$-$6) at all in spite of the strong detections with
single-dish observations. To better understand our SMA results, we
also carried out single-dish observations in CS (7$-$6) and
C$^{34}$S (7$-$6) with the 10 m Heinrich Hertz Submillimeter
Telescope (SMT) on Mount Graham, Arizona. The observations were
performed in the beam-switching mode. The primary beam is about
22\arcsec, and the spectral resolution is 1 MHz (corresponding to a
velocity resolution of $\sim$0.9 km s$^{-1}$). The temperature scale
\emph{T}$_A^*$ was obtained using standard vane calibration, and the
main beam temperatures (\emph{T}$_{mb}$) were derived through
\emph{T}$_{mb}$ = \emph{T}$_A^*$/$\eta$$_{mb}$, with the main beam
efficiency $\eta$$_{mb}$ = 0.5 \citep{nar05}. The rms noise level is
110 mK and 38 mK, respectively, for CS(7$-$6) and C$^{34}$S (7$-$6)
data in main beam brightness temperature units. We reduced the data
using the CLASS package.

\begin{figure*}

\centering
 \vspace{-6.7cm}
 \hspace{-1.4cm}
 \epsscale{1.05}
 \hspace{-.3cm}
 \plotone{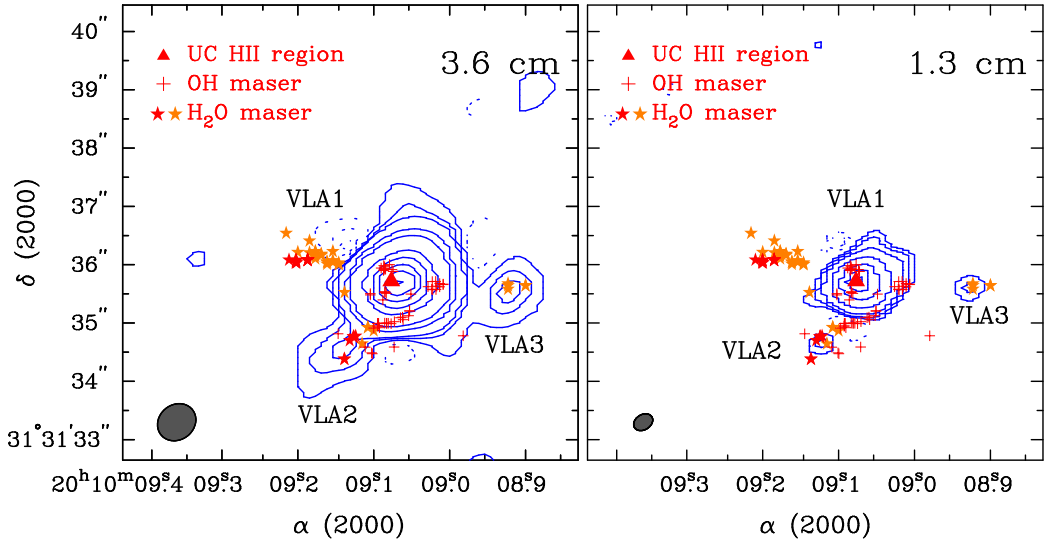}
 \vspace{-6.8cm}
\caption{Contour plots of continuum emission imaged toward ON~1
region at 3.6 cm (\emph{left}) and 1.3 cm (\emph{right}). Contour
levels are $-$15, $-$9, $-$3, 3, 9, 15, 50, 100, 400, 700, 1100,
1500, and 1900 $\times$ 0.05 mJy beam$^{-1}$ in the left panel, and
$-$7, $-$3, 3, 7, 20, 100, 400, 800, and 1200 $\times$ 0.06 mJy
beam$^{-1}$ in the right panel. Negative contours are dotted. In
each panel, the triangle represents the position of the UC \htwo
region ON~1 from \citet{kur94}, and the crosses mark the positions
of OH masers identified by \citet{nam06} and \citet{fis07a}. The red
and orange stars represent the positions of the water masers
reported by \citet{nag07} and \citet{tre08} respectively. The dark
ellipse denotes the synthesized beam. \label{fig-cont-vla}}
\end{figure*}

\begin{figure*}

\centering
 \vspace{-6.7cm}
 \hspace{-1.4cm}
 \epsscale{1.05}
 \hspace{-.3cm}
 \plotone{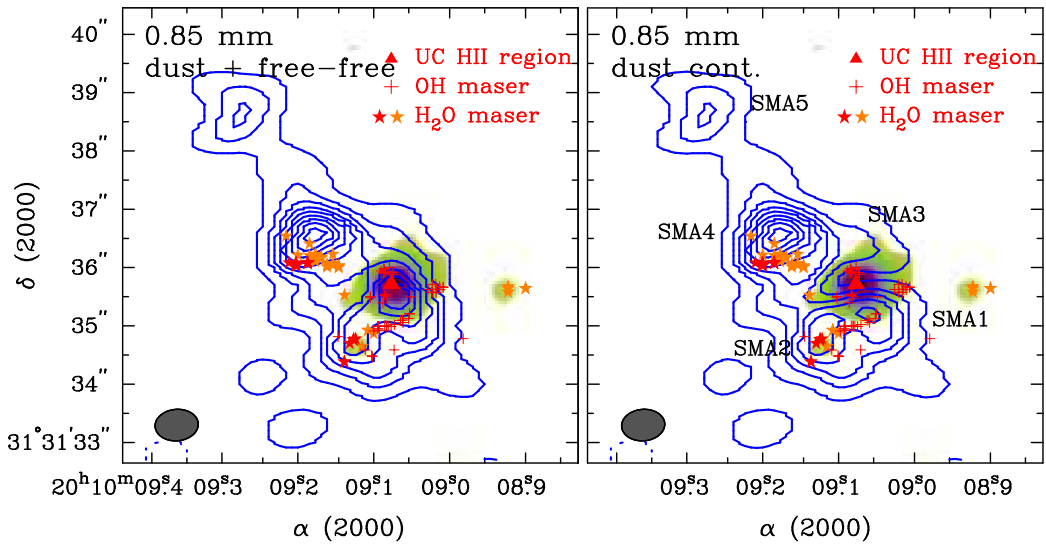}
 \vspace{-6.8cm}
\caption{\emph{left panel}: Contour plots of 0.85 mm continuum
imaged toward ON~1 region with the SMA. Contour levels are $-$9,
from 9 to 153 in step of 18 mJy beam$^{-1}$. Negative contours are
dotted. \emph{right panel}: 0.85 mm continuum with free-free
emission subtracted, see \S\ref{resultcont-sma} for the details. In
each panel, the overlaid color image is the VLA 1.3-cm continuum
shown in Figure~\ref{fig-cont-vla} in logarithmic scale. Caption as
in Figure~\ref{fig-cont-vla}. \label{fig-cont-sma}}
\end{figure*}

\section{Results \label{results}}
\subsection{3.6 cm and 1.3 cm Continuum \label{resultcont-vla}}
Figure~\ref{fig-cont-vla} shows the continuum maps at 3.6 cm
(\emph{left} panel) and 1.3 cm (\emph{right} panel) toward ON~1
region imaged with the VLA. At both wavelengths, three cm-wave
sources are detected within a field of $\sim$3\arcsec. The strongest
cm-wave source, denoted as VLA1, is identical to the previously
identified UC \htwo region ON~1, with its measured flux densities at
both 3.6 and 1.3 cm in close agreement with previous results
\citep[e.g.,][]{zhe85,kur94}. The spectral index of $\sim$0.4
inferred from the integrated 3.6 and 1.3 cm emission is consistent
with that being free-free emission with a turn-over frequency at
around 10 GHz \citep{zhe85}. To the southeast and west of the ON~1,
two additional cm-wave continuum sources, denoted as VLA2 and VLA3,
are located. VLA2 is the first time to be identified, while VLA3 is
detected by \citet{arg00} without special notification. Both VLA2
and VLA3 appear to be unresolved and are significantly weaker than
ON~1. In the same way, we deduced a spectral index of $-$0.7 for
VLA2 and $-$0.2 for VLA3. Although the spectral index of $-$0.7 is
similar to that of an extragalactic source, VLA2 is still most
likely related to the ON~1 region because VLA2 is also associated
with several star formation signatures as presented in
\S\ref{discuss-nature}. The positions, integrated flux densities,
and deconvolved source sizes of these three cm-wave sources are
summarized in Table~\ref{tb-cont}.

All the three cm-wave sources demonstrate maser activity. As shown
in Figure~\ref{fig-cont-vla}, VLA1 and VLA2 are associated with OH
maser spots, while VLA2 also harbors water maser emission
\citep{fis07a,nam06,nag07,tre08}. Furthermore, recent VLA
observations detected a group of water maser spots toward VLA3
\citep{tre08}. Therefore in addition to the UC \htwo region ON~1,
VLA2 and VLA3 may represent another two star forming sites in this
region. This is especially true for the case of VLA2 due to its
association of several other star formation signatures (see
\S\ref{resultcont-sma} and \S\ref{resultline}). A detailed
discussion about the nature of these cm-wave sources will be
presented in \S\ref{discuss-nature}.

\begin{deluxetable*}{lcccrrc}
 \tabletypesize{\scriptsize} \tablecaption{\sc Parameters of Continuum Sources}
 \singlespace
 \tablewidth{0pt}
 \tablehead{\colhead{} & \colhead{} & \multicolumn{2}{c}{Position(2000)}  & \colhead{S$_{int.}^a$}
 & \colhead{deconvolved size$^a$} & \colhead{Mass} \\
            \colhead{Source} &
            \colhead{wavelength} &
            \colhead{$\alpha$}  &
            \colhead{$\delta$} &
            \colhead{(mJy)} &
            \colhead{maj\arcsec min$\arcsec$ P.A.$^\circ$} &
            \colhead{(\emph{M}$_\sun$)}}
 \startdata
VLA results   &                &           &                   &  \\
~~VLA1   & 3.6cm  & 20 10 09.07   &  +31 31 35.7   &   136.4    &  0.50 0.35 $-$77.7&  \\
        & 1.3cm  & 20 10 09.07   &  +31 31 35.7   &   200.0    &  0.42 0.37 +73.4  &  \\
~~VLA2   & 3.6cm  & 20 10 09.16   &  +31 31 34.5   &     1.2    &  point source     &  \\
        & 1.3cm  & 20 10 09.12   &  +31 31 34.6   &     0.6    &  point source     &  \\
~~VLA3   & 3.6cm  & 20 10 08.92   &  +31 31 35.5   &     1.0    &  point source     &  \\
        & 1.3cm  & 20 10 08.93   &  +31 31 35.6   &     0.8    &  0.31 0.21 +56.1  &  \\
 \\
SMA results   &                  &                &            &                   &   \\
~~SMA1 & 0.85mm &  20 10 09.06  &  +31 31 35.0   &   320     &  1.23 1.16 $-$16.1 & 4.4 \\
~~SMA2 & 0.85mm &  20 10 09.13  &  +31 31 34.8   &   190     &  0.96 0.79 +29.8   & 2.6 \\
~~SMA3 & 0.85mm &  20 10 09.07  &  +31 31 36.5   &    60     &  0.86 0.26 $-$24.1 & 0.8 \\
~~SMA4 & 0.85mm &  20 10 09.18  &  +31 31 36.5   &   470     &  0.99 0.82 ~+0.5   & 6.4  \\
~~SMA5 & 0.85mm &  20 10 09.28  &  +31 31 38.6   &   170     &  1.32 1.00 +22.0   & 2.3  \\

 \enddata
 \tablenotetext{a}{derived from 2-D Gaussian fits \label{tb-cont}}
\end{deluxetable*}

\subsection{0.85 mm Continuum \label{resultcont-sma}}
The \emph{left} panel of Figure~\ref{fig-cont-sma} shows the 345 GHz
(0.85 mm) continuum emission toward ON~1 region imaged with the SMA.
Within a field of $\sim$5$\arcsec$ (i.e., 0.05 pc at 1.8 kpc),
several submillimeter continuum features can be discerned. The two
strongest peaks have been reported in our previous snapshot
observations \citep{su04}, while the weaker one is associated with
the UC \htwo region ON~1. Since 0.85 mm continuum could arise from
either dust or ionized gas, we estimate the dust contribution to the
observed 0.85 mm flux density. First, we extrapolate the free-free
emission as measured at cm wavelengths to 0.85 mm. As mentioned in
\S\ref{resultcont-vla}, the turn-over frequency of the UC \htwo
region ON~1 is around $\sim$10 GHz, indicating a flux density ratio
of $\sim$0.76 between 0.85 mm and 1.3 cm calculating from a
optically thin spectral index of $-$0.1. Secondly, we convolve the
1.3 cm map to a Gaussian beam of 0.74$\arcsec$$\times$0.54$\arcsec$
to match the angular resolution of the 0.85 mm map and regrid the
convolved 1.3 cm map to have the same coordinate system as that of
the 0.85 mm map. We then apply uniformly a scaling factor of 0.76 to
all pixels in the regrided 1.3 cm image to obtain an extrapolated
free-free image at 0.85 mm. Finally the latter image is subtracted
from the observed 0.85 mm map in order to obtain a continuum map
without contribution of free-free emission, as shown in the
\emph{right} panel of Figure \ref{fig-cont-sma}.

The 0.85 mm continuum shown in the \emph{right} panel of Figure
\ref{fig-cont-sma} is then presumably produced solely by dust in
spite of the scaling factor 0.76 derived only from the UC \htwo
region ON~1 (i.e., VLA1). The reasons are first VLA2 is weaker and
has negative spectral index (inferred from 3.6 and 1.3 cm), thus its
extrapolated ionized gas component at 0.85 mm should be negligible.
Secondly, we can show that the other sub-mm features can not be
hypercompact (HC) \htwo regions, since the turnover frequency of HC
\htwo regions is unlikely higher than $\sim$100 GHz. For example,
for a HC \htwo region with an emission measure of
1$\times$10$^{10}$, among the highest known value
\citep[e.g.,][]{dep97}, the turn-over frequency is $\sim$50 GHz
only. Thus for any HC \htwo region weaker than 0.18 mJy (i.e.,
$\lesssim$3$\sigma$ rms noise levels) at 22 GHz, its extrapolated
free-free emission at 0.85 mm is also negligible, i.e., $\lesssim$ 1
mJy.

The major difference between the two panels in
Figure~\ref{fig-cont-sma} is caused by the contribution of free-free
emission at 0.85 mm from the \htwo region ON~1. As shown in the
\emph{right} panel of Figure \ref{fig-cont-sma}, after removing
free-free contribution, the 0.85 mm dust image clearly exhibits a
cavity surrounding the UC \htwo region ON~1, and in total five dust
sources (denoted as SMA1, 2, 3, 4, \& 5) are further identified. The
original sub-mm peak toward the \htwo region ON~1 is split into two
sources (i.e., SMA1 \& SMA3). Morphologically speaking, SMA3 can be
discerned as the extended feature associated with SMA4. This is,
however, unlikely the case because of the different molecular
signatures between SMA3 and SMA4 (\S\ref{resultline}). The nature of
these sub-mm dust sources will be discussed in \S
\ref{discuss-nature}. We fit the 0.85 mm dust image with
2-dimensional Gaussian structures to obtain the position, integrated
flux and deconvolved size of each source. The results as well as the
inferred dust and gas masses (see below) are listed in
Table~\ref{tb-cont}. Assuming a gas-to-dust ratio of 100, a
(uniform) dust temperature of 30 K, a dust opacity $\kappa$$_{\nu}$
= 0.006 ($\nu$/245 GHz)$^{\beta}$ cm$^{2}$ g$^{-1}$
\citep{kra98,she02}, and an opacity index $\beta$ = 1.5, we estimate
the dust and gas masses of the sub-mm sources are in the range of
0.8 to 6.4 M$_\odot$.

\begin{figure*}
\centering
 \vspace{-6.7cm}
 \hspace{-1.4cm}
 \epsscale{1.05}
 \hspace{-.3cm}
 \plotone{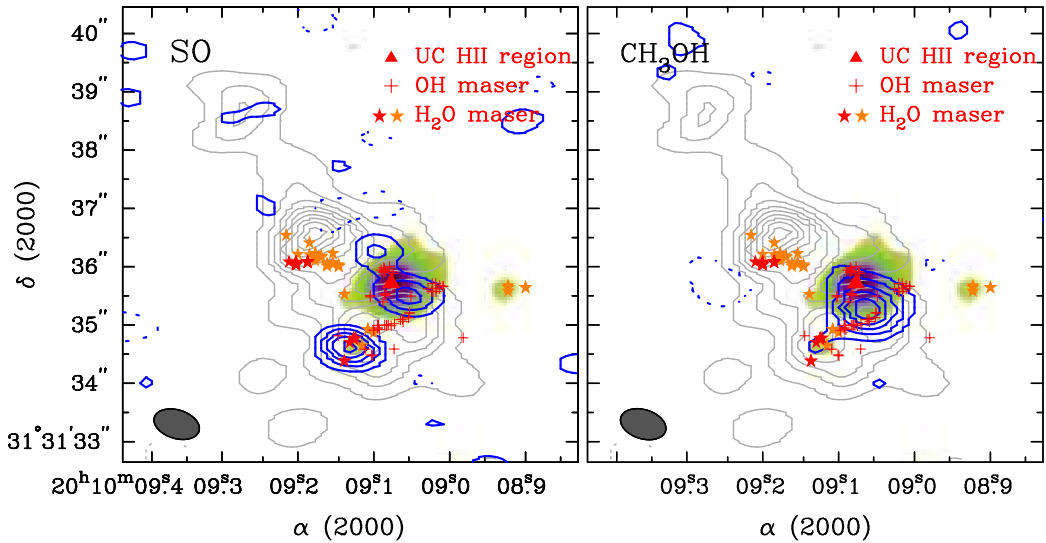}
 \vspace{-6.8cm}
 \caption{Contour plots of the integrated SO (8$_8$$-$7$_7$) (\emph{left}) and
 CH$_3$OH (13$_1$A$^-$$-$13$_0$A$^+$)
(\emph{right}) emission toward ON~1 region.  Contour levels in the
\emph{left} panel are $-$1.2 and from 1.2 to 9.6 in steps of 1.2 Jy
beam km s$^{-1}$, and in the \emph{right} panel are $-$1.05 and from
1.05 to 8.4 in steps of 1.05 Jy beam km s$^{-1}$. The 0.85 mm dust
continuum shown in Figure \ref{fig-cont-sma} \emph{right} is
overlaid as light contours.  The overlaid color image is the VLA
1.3-cm continuum shown in Figure~\ref{fig-cont-vla} in logarithmic
scale. Caption as in Figure~\ref{fig-cont-vla}. \label{fig-line}}
\end{figure*}

\begin{figure}
 \vspace{-.3cm}
 \hspace{.1cm}
 \includegraphics[angle=270,scale=.38]{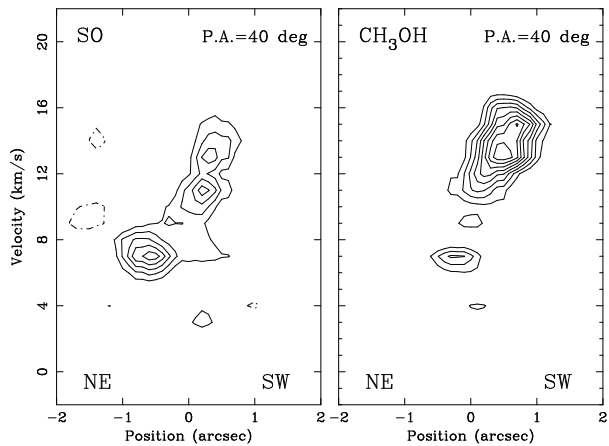}
 \vspace{-.7cm}
 \caption{Position-velocity diagram of the
 SO (8$_8$$-$7$_7$) (\emph{left}) and CH$_3$OH (13$_1$A$^-$$-$13$_0$A$^+$) (\emph{right})
 along the position angle of 40$^\circ$. Contour levels are $-$0.24, and from 0.24 to 0.88 in steps of 0.16 Jy beam$^{-1}$
 in the left panel, and from 0.24 to 1.36 in steps of 0.16 Jy beam$^{-1}$ in the right
 panel.
 \label{soch3ohpv}}
\end{figure}

\subsection{Morphology and Kinematics of the SO \& CH$_3$OH Lines \label{resultline}}
The \emph{left} panel of Figure \ref{fig-line} shows the integrated
SO (8$_8-$7$_7$) emission detected toward ON~1 region. The SO
emission can be detected from 6 to 15 km s$^{-1}$ approximately, and
is mainly concentrated around the positions of VLA1 and VLA2. As
shown in the \emph{left} panel of Figures \ref{fig-line} \&
\ref{soch3ohpv}, the SO line toward VLA1 (i.e., ON~1) traces an
elongated structure along a position angle of $\sim$40$^\circ$, with
blueshifted gas located to the northeast and redshifted to the
southwest. Furthermore, there appears to be a SO cavity surrounding
VLA1, i.e., the SO emission at the position of VLA1 is significantly
weaker than the northeast and southwest lobes. The strongest SO peak
spatially coincides with VLA2 and SMA2 very well, and is associated
with OH and H$_2$O masers. No clear velocity gradient of the SO line
toward VLA2 can be discerned. The \emph{right} panel of
Figure~\ref{fig-line} shows the integrated emission of CH$_3$OH
(13$_1$A$^-$$-$13$_0$A$^+$) detected toward ON~1 region. The
CH$_3$OH emission is discerned from 7 to 16 km s$^{-1}$ and is
mainly concentrated in the close vicinity of VLA1. In addition to
the main source, there appears to be weak CH$_3$OH emission
associated with VLA2.

As shown in the \emph{right} panel of Figure \ref{soch3ohpv}, the
kinematic structure of the main methanol source is similar to that
seen in SO line, although a spatial differentiation between the two
species is clearly discerned. Most of methanol gas is located to the
southwest of the UC \htwo region ON~1, while the integrated SO
emissions in the northeast and southwest of ON~1 are comparable. The
peak positions of the southwest source seen in SO and CH$_3$OH are
also slightly different, with an offset of $\sim$0.2$\arcsec$. Since
the upper energy levels of the two transitions are fairly different
($\emph{E$_u$}$ of SO (8$_8-$7$_7$) = 87.4 K and $\emph{E$_u$}$ of
CH$_3$OH (13$_1$A$^-$$-$13$_0$A$^+$) = 227.6 K), the distinct
contrast in SO and CH$_3$OH can result from different physical
conditions in this region. Interestingly, although the SMA4 shows
strong dust continuum, neither SO nor CH$_3$OH emission is detected
toward this sub-mm dust source.

\section{Discussion \label{diss}}

\subsection{Nature of Continuum \& Molecular-Line Sources \label{discuss-nature}}

It has been known that high-mass stars tend to form in groups.
Indeed, our results of the VLA and SMA observations clearly
demonstrate a group of continuum sources in this region. In addition
to the known UC \htwo region, which is powered by a B0 star, we
identify here two more cm-wave sources and five more sub-mm sources.
Below we discuss the nature of these continuum sources.

SMA2/VLA2 exhibits several star formation signposts such as cm-wave
emission (i.e., VLA2), H$_2$O and OH maser activities, and SO and
CH$_3$OH emission, and hence marks a newly identified star-forming
core toward ON~1 region. Assuming VLA2 to be a radio jet, we could
extrapolate the correlation of 3.6 cm radio continuum luminosity and
far-IR bolometric luminosity deduced from a sample of radio jets
associated with YSOs with luminosities $\lesssim$
10$^3$~\emph{L}$_\sun$ \citep{ang95,har02}, and infer a bolometric
luminosity of $\sim$6000~\emph{L}$_\sun$ for the young stellar
object (YSO) powering VLA2. In contrast, assuming VLA2 to be an UC
\htwo region, we could estimate the bolometric luminosity of \htwo
region powering source to be $\sim$2000~\emph{L}$_\sun$.
Furthermore, the dust and gas mass of the associated sub-mm
continuum source (i.e., SMA2) is $\sim$2.6 \emph{M}$_\sun$.
Therefore the available evidences suggest that SMA2/VLA2 likely
represents an intermediate-mass (or even high-mass) star forming
core.

The nature of VLA3, which is associated with water maser spots, is
less clear. In the same way, we estimate a bolometric luminosity of
$\sim$5000 \emph{L}$_\sun$ and $\sim$2000 \emph{L}$_\sun$,
respectively, for the VLA3 powering source by assuming VLA3 to be a
jet and an UC \htwo region. The non-detection of sub-mm continuum
toward VLA3 indicates an upper-limit to its gas mass of $\sim$0.13
\emph{M}$_\sun$ only, while the bolometric luminosity of
$\sim$2000$-$5000 \emph{L}$_\sun$ argues that the dust condensation
surrounding VLA3 is unlikely to be missed by our SMA observations.
In fact, the velocities of water maser spots associated with VLA3
span from $-$20 km~s$^{-1}$ to $-$9 km~s$^{-1}$, approximately 25
km~s$^{-1}$ shifted from the systemic velocity of the ON~1 region.
Thus probably VLA3 and its associated water maser spots trace a
blue-shifted outflow powered by a YSO within the ON~1 region.

The strongest sub-mm source, SMA4, is not associated with any star
formation signatures. Although \citet{nag07} proposed the highly
collimated water maser jet, located $\sim$0.8$\arcsec$ southwest of
the SMA4, to be associated with the SMA4, we argue that this is
unlikely the case, as described in \S\ref{discuss-agb}. Similarly,
SMA5 does not exhibit any signposts of star formation either. One
possible reason is that both SMA4 and SMA5 mark star formation
regions at very early evolutionary stages.

In \S\ref{resultcont-sma}, we identified two dust sources, SMA1 \&
SMA3, located to the north and southwest of the UC \htwo region
ON~1, after subtracting free-free emission from the 0.85-mm image.
The main methanol source coincides spatially with the SMA1 fairly
well. The SO emission to the southwest of the ON~1 is also likely
associated with the SMA1, and the SO gas to the northeast of the
ON~1 is likely associated with the SMA3. Kinematically speaking, the
velocity gradient of H$^{13}$CO$^+$ (1$-$0) along the
northeast-southwest direction toward the \htwo region ON~1 has been
reported by various authors, and is interpreted as either molecular
outflows \citep{kum04} or a rotating dense core \citep{lim02}. A
similar kinematics, which can be caused by molecular outflow,
rotation of dense gas, and expansion of \htwo region, is also seen
in SO and CH$_3$OH lines with our SMA observations. Below, we
re-examine these interpretations based on our high-spatial
resolution results and discuss the nature of SMA1 \& SMA3.

First, if the velocity structures seen in SO and CH$_3$OH result
from molecular outflow, the most promising outflow driving source
shown in our SMA subarcsecond images is the same as the young star
powering the UC \htwo region ON~1. In this scenario, one will expect
to identify signature of the ionized shell broken by the outflow
activities. Such a signature, however, can not be discerned from the
cm-wave images shown in Figure~\ref{fig-cont-vla}. Our results
therefore do not appear to favor the interpretation of molecular
outflow. The remaining two possibilities (i.e., expansion and
rotation) are not very easy to discriminate. With proper motion
measurements, \citet{fis07b} conclude that the motions of OH masers
are consistent with the expansion of the UC \htwo region ON~1. Since
both CH$_3$OH and SO emissions span similar velocity ranges with
those measured in OH masers, and the kinematic structures of OH
masers and CH$_3$OH and SO gas are similar, we suggest that the SO
and CH$_3$OH gas associated with SMA1 and SMA3 trace the expansion
of the \htwo region.

\subsection{Clustered Mode of Star Formation \label{discuss-cluster}}

Similar to many other massive star formation regions, the ON~1
region is also associated with an embedded star cluster identified
at near-infrared wavelengths \citep{kum02,kum03}. It has been argued
that in the Milky Way about 70\%$-$90\% stars are formed in
clusters, and several fundamental properties of the Galactic stellar
population, such as the initial mass function (IMF), are forged in
embedded star clusters \citep[for a recent review, see][]{lad03}.
The star formation efficiency (SFE, =
\emph{M$_{stars}$}/(\emph{M$_{gas}$}+\emph{M$_{stars}$})) is a
fundamental parameter of star and cluster formation processes, and
appears to be an indicator for the evolution of an embedded cluster.
Observations of nearby embedded clusters show that SFEs range from
about 10\% to 30\%, with the least evolved clusters having the
lowest SFEs \citep[see][and reference therein]{lad03}. Below, we
estimate the SFE in the ON~1 region.

The mass of the molecular clump associated with ON~1 have been
evaluated to be $\sim$300$-$400 \emph{M}$_\sun$ (scaled to a
distance of 1.8 kpc) from single-dish observations
\citep{mul02,shi03}. In contrast, no mass estimation of the
associated embedded cluster has been done. Actually, an accurate
estimation of stellar mass within an embedded cluster is not easy.
We roughly evaluate the cluster mass from the 2MASS Point Source
Catalog. We first count the number of the stars detected within a
radius of 60$\arcsec$ (approximately a factor of two larger than the
size of the molecular clump) from the central \htwo region ON~1, and
in total 58 stars are found. Assuming a mean mass of 0.5
\emph{M}$_\sun$ inferred from the general IMF \citep{kro02}, the
mass of the embedded cluster is 29 \emph{M}$_\sun$. We further
include the mass of the UC \htwo region powering star, which is
likely a 10 \emph{M}$_\sun$ B0-type star, and estimate a SFE of
$\sim$10\% in the ON~1 region. The relatively low SFE appears to
suggest star formation in the ON~1 region will continue.

Indeed, the identification of multiple centimeter and millimeter
continuum sources toward ON~1 region indicates an ongoing
(proto)cluster formation. Given the small amount of mass (i.e.,
$\sim$16 \emph{M}$_\sun$) in the sub-mm sources reported in
\S\ref{resultcont-sma}, nevertheless, the SFE in ON~1 region is
still far from 30\%. We propose three possibilities for the ON~1
region if it would attain a high SFE of 30\% as deduced for the more
evolved young clusters. First is that there will be new cores and
then stars forming later, second is that a significant fraction of
the clump mass has to be accreted onto the existing dust sources,
and/or third that a significant number of dust cores have been
missed by our SMA observations. For the second possibility, since
cores keep accreting gas from clump, estimating final star mass from
current core mass is difficult. Thus one may expect a non-correlated
relationship between star mass and core mass functions, as predicted
by the competitive accretion model \citep{bon04}.

Due to the lack of sufficient dynamical range, a proper
identification of dust continuum sources with masses $\lesssim$0.4
\emph{M}$_\sun$ can not be achieved by our SMA observations. That
most sub-mm dust sources detected toward ON~1 region are massive
than 2 \emph{M}$_\sun$, however, appears to suggest an absence of
low-mass (\emph{M} $\lesssim$ 2 \emph{M}$_\sun$) dust sources within
field of view. If the core mass function exists at such a star
formation scale, the lack of low-mass dust sources toward the
cluster center infers a mass segregation, i.e., massive cores
preferentially located near cluster center, in the ON~1 region. Such
core mass segregation is probably related to the stellar mass
segregation often seen in young clusters \citep[e.g.,][and reference
therein]{zin07}.

\begin{figure}
 \vspace{-.9cm}
 \epsscale{1.1}
 \plotone{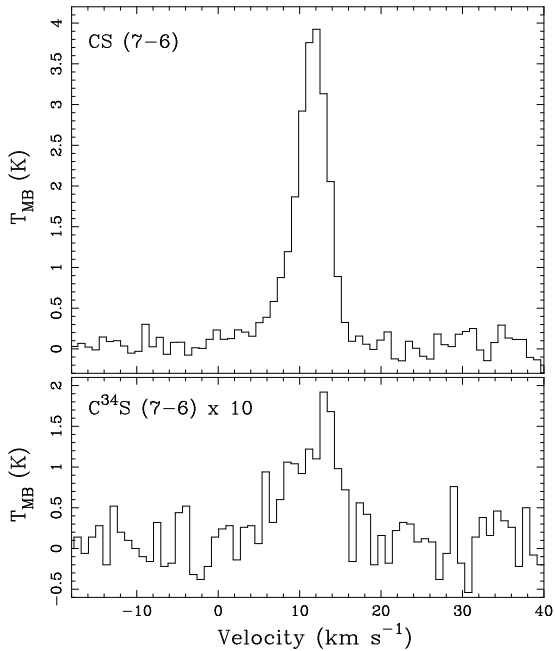}
 \vspace{-1.1cm}
\caption{Spectra of CS (7$-$6) (\emph{upper}) and C$^{34}$S (7$-$6)
(\emph{lower}) toward the Onsala~1 region observed with the 10-m
Heinrich Hertz Submillimeter Telescope at about 22$\arcsec$ beam.
For clarity, the C$^{34}$S data have been multiplied by a factor of
10. \label{fig-cs-hht}}
\end{figure}

\subsection{Non-detections of CS (7$-$6) and HCN (4$-$3)  \label{discuss-cs}}

Although CS (7$-$6) gas has been detected toward ON~1 region with a
peak flux density of $\sim$120 Jy in a 20$\arcsec$ beam using
single-dish telescopes \citep{plu92}, it is not detected at all with
our SMA observation in the image plane at a rms noise level of
$\sim$100 mJy. To estimate the column density upper limit of the CS
gas, we assume the CS (7$-$6) to be optically thin and in LTE and
adopt a gas temperature of 30 K, the same as that of the sub-mm
(dust) continuum emission. We derive a CS column density of
$\lesssim$ 9$\times$10$^{13}$. Together with the total (gas) column
density of (1.1$-$6.0)$\times$10$^{24}$ derived from the five 0.85
mm (dust) sources, we estimate a CS abundance \emph{X}(CS) of
$\lesssim$ (8$-$2)$\times$10$^{-11}$, with the highest value (i.e.,
8$\times$10$^{-11}$) for the SMA5 and the lowest for the SMA4. The
inferred CS abundance will be increased if the adopted gas
temperature is higher. For example, the upper-limit CS abundance
will become (11$-$2)$\times$10$^{-10}$ if assuming a gas temperature
of 300 K. We conclude that the inferred CS abundance, at least for
the more massive sub-mm sources such as SMA4, is likely much smaller
than the typical CS abundance of 1$\times$10$^{-9}$ toward high-mass
star-forming cores obtained from single-dish observations
\citep{shi03}. With the same assumptions, similarly we estimate the
HCN column density being $\lesssim$ 2.0$\times$10$^{13}$ and the
abundance \emph{X}(HCN) $\lesssim$ (0.3$-$2)$\times$10$^{-11}$. In
star forming cores, the typical \emph{X}(HCN) is about a few
$\times$10$^{-9}$ \citep{jor04}.

The direct interpretation of the non-detection of CS (7$-$6) and low
CS abundance is that there is indeed a depletion of CS in the inner
region of the massive star forming core associated with ON~1, and
the majority of the CS (7$-$6) emission detected by single-dish
telescope should arise from an extended envelope. A proper
explanation for the CS depletion is not easy. \citet{taf02} reported
a CS depletion at the phase of starless cores, which might be able
to explain the non-detection of CS (7$-$6) gas at some dust
condensations in the ON~1 region. However, ON~1 region appears to
harbor (some) YSOs which are more evolved than the pre-stellar-core
phase. For example, SMA2 is associated with masers of H$_2$O and OH
and cm-wave continuum, and SMA1 exhibits strong CH$_3$OH emission
with $\emph{E$_u$}$$\sim$300 K. Both SMA1 \& SMA2 are most likely
evolved than the phase of pre-stellar core and their surrounding gas
should be hot enough to evaporate CS from grain mantles to gas
phase.

An alternative situation is that the CS (7$-$6) emission from the
outer envelope is optically thick, quite smoothly distributed and
resolved out, one therefore cannot see any embedded compact features
spanning the same velocities. This is, however, unlikely the case
for the ON~1 region. As shown in Figure \ref{fig-cs-hht}, our
single-dish observations with the 10-m SMT indicate that the CS
(7$-$6) emission toward ON~1 region, with a main-beam temperature of
$\sim$4 \emph{K}, is far from optically thick. Moreover, that the
line ratio of CS (7$-$6) and C$^{34}$S (7$-$6) is very close to the
$^{32}$S/$^{34}$S abundance ratio of $\sim$22 \citep{wil94} also
argues the optically thin CS (7$-$6) emission toward ON~1 region.

Although it is unlikely that the CS (7-6) emission is optically
thick over the whole single-dish beam of $\sim$22$\arcsec$, we can
not rule out the possibility that the CS (7$-$6) of $\sim$50$-$100 K
is opaque within a region of $\sim$5$\arcsec$ or so toward the ON~1.
Such conditions can still account for the non-detection of the CS
(7$-$6) in our SMA observations. Interferometric observations at
3\arcsec$-$5\arcsec resolution, which can provide information to
link the large-scale results from single-dish observations and the
small-scale measurements with our SMA observations presented here,
should help to address this issue.

\subsection{Nature of the Extremely High-Velocity Water Masers \label{discuss-agb}}

One of the most dramatic results from maser observations toward ON~1
region is the detection of extremely high-velocity H$_2$O maser
spots, with velocity $\sim$ $V_{lsr}$ $\pm 60$ km s$^{-1}$, located
approximately $\sim$1.5$\arcsec$~ to the east of the UC \htwo region
ON~1 \citep{nag07}. Such a velocity coverage is among the broadest
range of H$_2$O maser emission detected in (high-mass) star forming
regions. In particular these extremely high-velocity masers spread
into two small ($\sim$15 $\times$ 15 mas) regions separated
approximated 0.2$\arcsec$ along the east-west direction and display
a clear velocity gradient, with blue spots to the east and red to
the west \citep{nag07}.

Together with proper motion measurements, \citet{nag07} suggest that
these high-velocity H$_2$O masers trace an outflow aligned roughly
in the east-west direction driven by a YSO in this region. The
collimated jet seen in water masers, however, is not centered on any
sub-mm (dust) peak or molecular gas condensation traced by SO and
CH$_3$OH. With the existence of a power-law correlation between
\emph{L}$_{H_2O}$ and \emph{L}$_{FIR}$ \citep{pal93}, the measured
high water maser luminosity \citep{nag07} indicates a pumping source
of luminous (\emph{L}$_{bol}\approx$ 2$\times$10$^4$
\emph{L}$_\sun$) and massive YSO unlikely to be missed with our SMA
observations. To our knowledge, furthermore, such an extremely
high-velocity water maser system exhibiting a compact
($\sim$300$-$400 AU) jet-like structure has never been detected in
other star forming region.

Alternatively, stars at asymptotic giant branch (AGB) stage can
demonstrate maser activities of various species (i.e., H$_2$O, OH,
and SiO) \citep[][and reference therein]{dea07, ima07}. The
velocities of masers detected in evolved stars usually span less
than 40 km s$^{-1}$, but extremely high-velocity ($\sim$100 km
s$^{-1}$) water masers have been detected toward approximately 10
evolved stars \citep[for a recent review, see][]{ima07b}. These
sources are the so-called ``water-fountain'' sources. They exhibit
jet-like signatures in both kinematics and morphology, with extents
of a few $\times$ 100 AU and dynamical timescales of a few tens
years. The origin of the high-velocity collimated jets around such
AGB stars is not well understood. One possible scenario is that the
collimated jets are shaped by stellar magnetic field \citep{vel06}.

Given the lack of identification of a young star to be responsible
to power the water masers and the similarities in both morphology
and kinematics of the H$_2$O maser spots to the so-called
``water-fountain'' sources, we suspect that the collimated water jet
toward ON~1 region may be driven by an evolved star. If this is the
case, it is interesting to know whether the water fountain jet as
well as its driving source are really located within the massive
star forming region ON~1 or just along the line of sight. A precise
estimate of the latter probability can not be done due to the lack
of the sample statistics of water fountain jets. We consider that
the probability of the latter is very small. Given a similar
systematic velocity of the ON~1 region and the collimated water jet,
a direct discriminate to this issue is not easy. A precise
determination of the distance to the water fountain jet and the ON~1
region by the annual parallax measurements with the VLBI will help
to answer this question.

\section{Summary}

We present results of interferometric observations toward the
high-mass star formation region ON~1 made with the SMA and VLA in
continuum and molecular lines at subarcsecond resolutions. Our main
results can be summarized as follows.

1. Observations with the VLA at both 1.3 cm and 3.6 cm identified
two new cm-wave continuum sources in addition to the known UC \htwo
region ON~1. All the three cm-wave sources demonstrate maser
activity.

2. Our SMA observations identified five sub-mm dust sources within a
field of $\sim$5$\arcsec$, indicating the multiplicity in the ON~1
region. The dust and gas masses of these sub-mm sources are in the
range of 0.8 to 6.4 \emph{M}$_\sun$.

3. SMA2 exhibits several star formation signatures such as cm-wave
emission (i.e., VLA2), H$_2$O and OH maser activities, as well as SO
and CH$_3$OH emission, and hence marks a newly identified
star-forming core in the ON~1 region. With an inferred far-IR
bolometric luminosity of $\sim$6000 \emph{L}$_\sun$, SMA2 represents
an intermediate-mass or even high-mass star forming core.

4. SMA1 and SMA3 are associated with SO and CH$_3$OH lines. The
kinematic structures of both SO and CH$_3$OH appear consistent with
the expansion of the UC \htwo region ON~1.

5. A low star formation efficiency of $\sim$10\% suggest that star
formation in the ON~1 region will continue. This is consistent with
the identification of multiple YSO candidates in the ON~1 region.

6. While strong CS (7$-$6) emission was detected toward the ON~1
region with single-dish observations, it is not detected at all by
our SMA observations. The results of single-dish observations in CS
(7$-$6) and C$^{34}$S (7$-$6) rule out the possibility that the CS
line was totally resolved out due to the smoothly distributed and
optically thick CS emission from the extended envelope over the
single-dish beam of 22$\arcsec$. The non-detection of CS (7$-$6)
emission may imply a depletion of CS in the center of the ON~1
region.

7. There is a lack of identification of a young star to be
responsible for powering the highly collimated high-velocity water
jet identified in the ON~1 region. We therefore suggest that these
masers, like H$_2$O maser spots in the so-called ``water-fountain''
sources, are driven by an evolved star.

\acknowledgments We thank all SMA staff for their help during these
observations. S.-Y. L. and Y.-N. S. thank the National Science
Council of Taiwan for support this work through grants NSC
97-2112-M-001-006-MY2.

\end{document}